\documentclass[%
 reprint,
preprintnumbers,
nofootinbib,
 amsmath,amssymb,
 aps,
prd,
 showkeys,
]{revtex4-2}

\usepackage{graphicx}% Include figure files
\usepackage{dcolumn}% Align table columns on decimal point
\usepackage{bm}% bold math
\usepackage{amsfonts}
\usepackage{enumitem}% want to avoid column break on page 3 at start of enumerate, using [beginpenalty=10000]

\newcommand{\Q}{\mathbb{Q}}

\newcommand{\F}{\mathbb{F}}
\newcommand{\bigO}{\mathcal{O}}

\begin{document}

\preprint{OUTP-23-14P}

\title{$p$-adic reconstruction of rational functions in multi-loop amplitudes}

\author{Herschel A. Chawdhry}
 \email{hchawdhry@fsu.edu}
\affiliation{%
Department of Physics, University of Oxford, Parks Road, OX1 3PU, United Kingdom\\
Department of Physics, Florida State University, 77 Chieftan Way, Tallahassee FL 32306, USA
}%

\date{\today}

\begin{abstract}
Numerical reconstruction techniques are widely employed in the calculation of multi-loop scattering amplitudes.
In recent years, it has been observed that the rational functions in multi-loop calculations greatly simplify under partial fractioning.
In this article, we present a technique to reconstruct rational functions directly in partial-fractioned form, by evaluating the functions at special integer points chosen for their properties under a $p$-adic metric.
As an application, we apply this technique to reconstruct the largest rational function in the integration-by-parts reduction of one of the rank-5 integrals appearing in 2-loop 5-point full-colour massless amplitude calculations in Quantum Chromodynamics (QCD).
The number of required numerical probes (per prime field) is found to be around 25 times smaller than in conventional techniques, and the obtained result is 130 times smaller.
The reconstructed result displays signs of additional structure that could be used to further reduce its size and the number of required probes.
\end{abstract}

 %Use showkeys class option if keyword display desired
\maketitle

\section{Introduction}\label{sec:intro}
Multi-loop scattering amplitudes are a cornerstone of high-precision predictions in particle physics.
These amplitudes are challenging to calculate and one of the key bottlenecks in these computations is the symbolic calculation of complicated multi-variable rational functions, which appear in the final expression for the amplitudes as well as related results such as tables of Integration-By-Parts identities (IBPs)~\cite{Tkachov:1981wb,Chetyrkin:1981qh}.

In principle, calculating these rational functions only requires elementary polynomial arithmetic: addition, subtraction, multiplication, and division. But the rational functions become large at intermediate stages of the calculation (as compared to the initial or final stages) and so the arithmetical operations at intermediate stages become very slow. This ``intermediate-expression swell'' phenomenon is well known in computer algebra, and in that field it has been common since the 1960s~\cite{Borosh1966ExactSO,collins1968computing,10.1145/321784.321787,10.1145/800206.806398,zurGathenGerhard} to instead perform such calculations numerically at sample points in a finite field $\F_p$ or $p$-adic field $\Q_p$ and then construct the analytic form of the final result by interpolation techniques.
The computational time of this approach is largely determined by the number of required numerical samples, which depends only on the complexity of the final expression and thereby bypasses the complexity of intermediate stages. During the last decade, finite-field methods have been directly adopted in multi-loop amplitude calculations with much success~\cite{vonManteuffel:2014ixa,Peraro:2016wsq,Klappert:2019emp,Peraro:2019svx,Klappert:2020aqs}.

The rational functions reconstructed by finite-field methods are typically interpolated in common-denominator form, i.e. as a ratio of two polynomials.
In multi-loop calculations in recent years, it has been observed that these rational functions, once reconstructed, can be simplified by up to 2 orders of magnitude by partial fractioning them~\cite{Abreu:2019odu,Boehm:2020ijp,Agarwal:2021grm,Agarwal:2021vdh,Bendle:2021ueg,Heller:2021qkz,Chawdhry:2021mkw,Badger:2022mrb}.
In principle, it would be desirable and advantageous to exploit this simplification earlier, i.e. \emph{during} the reconstruction, so as to reduce the number of probes required by up to 2 orders of magnitude.
But as we will show in section~\ref{sec:preliminary_remarks}, the simplification under partial fractioning is not a generic feature of rational functions, but instead seems to be a special property of the specific rational functions appearing in IBPs and amplitudes.
Exploiting this simplification therefore requires developing specialised techniques that go beyond those used in generic computer algebra calculations.
This is in contrast to the use of finite-field methods, which are the computer algebraist's standard solution to the widespread computer algebra phenomenon of intermediate-expression swell.

To date, some work has been performed with the aim of optimising numerical reconstruction methods for high-energy physics use cases.
As will be explained in sec.~\ref{sec:preliminary_remarks}, it is possible to guess~\cite{Abreu:2018zmy,Heller:2021qkz} the common denominator of the rational functions in multi-loop calculations, thereby reducing the number of required numerical probes by a factor of 2.
A partial-fractioned reconstruction technique based on very high-precision floating-point evaluations was presented in Refs~\cite{Laurentis:2019bjh,DeLaurentis:2020qle}.
Within a finite-field context, some benefits may also be obtained by reconstructing in one variable at a time and performing single-variable partial fractioning at some intermediate stages~\cite{10.1145/800206.806398,Bendle:2019csk,Badger:2021imn,Badger:2021nhg,Badger:2021ega,Abreu:2021asb}, possibly in conjunction with expanding in $\epsilon$, where $D=4-2\epsilon$ is the spacetime dimension variable.
Techniques based on algebraic geometry and evaluations in $\Q_p$ have been proposed~\cite{DeLaurentis:2022otd,DeLaurentis:2023nss,DeLaurentis:2023izi} for eliciting information about the numerator of a rational function prior to performing a finite-field reconstruction, and Ref.~\cite{Campbell:2022qpq} mentions combining these with the methods of Ref.~\cite{Laurentis:2019bjh}.

In this work, we present a new technique to reconstruct rational functions directly in partial-fractioned form.
Our technique uses $p$-adic probes to reconstruct the rational functions one partial-fractioned term at a time, exploiting the simplification under partial fractioning and exposing hints of further patterns and structure.
We will apply this technique to reconstruct one of the largest rational functions appearing in the table of 2-loop 5-point massless non-planar IBPs, a highly complicated example which is at the edge of the capabilities of current tools and methods.
It will be shown that for this example function, our technique requires 25 times fewer numeric ($\Q_p$) probes than conventional ($\F_p$-based) reconstruction, and leads to a 130-fold reduction in the size of the final result.

The rest of this article is organised as follows. In section~\ref{sec:preliminary_remarks}, we give preliminary remarks to motivate our strategy. In section~\ref{sec:padics} we briefly describe some properties of $\Q_p$ that we use. In section~\ref{sec:method} we present our reconstruction method. In section~\ref{sec:results} we analyse the results and performance. A summary and concluding remarks are presented in section~\ref{sec:conclusion}.

\section{Preliminary remarks}\label{sec:preliminary_remarks}

The focus of this work is on reducing the number of numerical probes required to reconstruct rational functions, since this is the dominant computational cost.
This number is independent of the choice of method for performing the numerical probes.
We therefore develop and test our techniques by taking a large rational function for which we have an analytic expression, and performing so-called ``black-box'' probes on it.
Specifically, we choose to work with the largest rational function appearing in the largest IBP expression used in Ref.~\cite{Agarwal:2021vdh} for calculating the complete set of full-colour 2-loop amplitudes for $pp \rightarrow \gamma \gamma j$ in massless Quantum Chromodynamics (QCD).
This is also the second-largest IBP expression needed for the corresponding 3-jet amplitudes of Ref.~\cite{Agarwal:2023suw}.
We will call this rational function $R_*$.

The inspiration for our method is the observation that, as shown in Table~\ref{tab:R_simplification_partial_fractioning}, $R_*$ becomes $\sim 100$ times smaller after multi-variate partial fractioning, compared to its size in common-denominator form.
\begin{table}
\caption{
\label{tab:R_simplification_partial_fractioning}
Simplification of $R_*$ under partial~fractioning.
Common-denominator form has numerator fully expanded and denominator fully factorised.
Partial-fractioned form is obtained using \textsc{MultivariateApart}~\cite{Heller:2021qkz} with option \texttt{UseFormProgram->True}.
(See Table~\ref{tab:results} for results obtained in this work.)
Sizes are as reported using \texttt{ByteCount} in \textsc{Mathematica}.
Number of free parameters is obtained by counting the number of terms in the fully-expanded numerator(s). 
}
\begin{tabular}{c|c|c}
Form of expression	&	Size	& Parameters to fit	\\
\hline
Common-denominator	& 605 MB &	1,369,559	\\
Partial-fractioned	& 4 MB	& 14,558
\end{tabular}
\end{table}
Thus, if everything were known about $R_*$ except the integer coefficients in the numerators of the partial-fractioned expression, only~14,558 free parameters would need to be fitted and so 94 times fewer numerical evaluations would be required compared to conventional reconstruction techniques.
The aim of this work is to achieve a speed-up of this nature \emph{without} requiring any prior knowledge about $R_*$.

The denominator of the common-denominator form of the rational function of interest is usually straight-forward to obtain because it is a product of simple factors, whose powers can be found by performing a numerical probe at an integer kinematic point where some of those factors are prime numbers~\cite{Heller:2021qkz}.
The factors themselves can be guessed in a variety of ways: either by examining the symbol (if known) of the differential equations describing the master integrals~\cite{Abreu:2018zmy}, or by examining the denominators of the rational functions in the solution to a low-numerator-rank IBP system, or simply by trial and error.

We believe, however, that the simplification in Table~\ref{tab:R_simplification_partial_fractioning} does not follow from the mere fact that the denominator of $R_*$ factorises.
To see this, we considered several rational functions from the solutions to the 2-loop 5-point massless IBP equations.
For each rational function $R$ we constructed a second rational function, $\tilde{R}$, obtained by taking $R$ in common-denominator form and replacing with random numbers all coefficients in its fully-expanded numerator, while leaving the denominator unchanged.
We observed that each $R$ gets simplified upon partial fractioning, and the simplification factor is largest for the largest rational functions.
Yet if we partial-fraction $\tilde{R}$, no simplification occurs; indeed the partial-fractioned form of $\tilde{R}$ is typically slightly larger than its common-denominator form, regardless of whether it is measured using \texttt{ByteCount} or the number of free numerator parameters.
We conclude that the above-mentioned simplification of $R_*$ upon partial fractioning does not occur for generic rational functions, but is instead a special property of $R_*$, which we conjecture will generalise to many IBP and amplitude expressions.\footnote{We emphasise that the selection of $R_*$ as working example was not on the basis of any such properties, but was on the contrary because it is an exceedingly complicated expression that is on the boundary of current computational techniques.} Therefore, it should be expected that exploiting this simplification will require custom techniques beyond those developed for generic computer algebra problems.

To understand the reason for the simplification in Table~\ref{tab:R_simplification_partial_fractioning} and guide a strategy for exploiting it, we applied \textsc{MultivariateApart}~\cite{Heller:2021qkz}\footnote{\textsc{MultivariateApart} implements the Leinartas algorithm~\cite{leinartas1978factorization}, which was also implemented in Refs.~\cite{raichev2012leinartass, Meyer:2016slj, Meyer:2017joq, Abreu:2019odu}.} to several examples of $R$ and $\tilde{R}$. In each case we compared the resulting expressions\footnote{In this work, we find it helpful to adopt a convention of allowing no overall integer factors in the denominator of a partial-fractioned form, instead preferring to put such factors into the numerator, e.g. preferring $\frac{\frac{y^2 z}{3} + \frac{8y z^2}{3}}{(x+y)(x-z)}$ instead of $\frac{y^2 z + 8y z^2}{3(x+y)(x-z)}$.}
\begin{equation}\label{eq:R_pfed}
R = \sum_i \frac{n_i}{d_i},
\end{equation}
\begin{equation}\label{eq:Rtilde_pfed}
\tilde{R} = \sum_j \frac{\tilde{n}_j}{d_j}.
\end{equation}
For all the examples studied, we observed that the sum in eq.~\eqref{eq:R_pfed} contains fewer terms than the sum in eq.~\eqref{eq:Rtilde_pfed}.
Furthermore, all of the terms in eq.~\eqref{eq:R_pfed} also appear in eq.~\eqref{eq:Rtilde_pfed}, albeit with different numerators---in other words, $\{d_i\}$ is a subset of $\{d_j\}$.
We noted that if the partial-fractioned terms that are present in $\tilde{R}$ but vanish in $R$ could be identified in advance, it would give a large simplification. In the case of $R_*$, we estimated this simplification would be a factor of 28 compared to the common-denominator form, reducing the number of free parameters from 1,369,559 to 48,512.
For this reason, the core aim of our strategy presented in sec.~\ref{sec:method} is to identify, as cheaply as possible, which partial-fractioned terms vanish.
The remaining factor of $\frac{48,512}{14,558} \approx 3.3$ between this and the figure in Table~\ref{tab:R_simplification_partial_fractioning} arises because many of the partial-fractioned terms in eq.~\eqref{eq:R_pfed} have numerators containing fewer terms than the most generic polynomial that could be expected; we will leave to future work the exploitation of this further simplification.\footnote{In contrast to the $\mathcal{O}(30-100)$ simplification factors targeted here, we note that performing single-variable partial fractioning on these examples of $R$ produces simplifications by only a factor of between 2 and 12 relative to common-denominator form. The precise factor depends on the choice of variable with respect to which such a single-variable partial-fractioning is performed} In addition, our results (see sec.~\ref{sec:results}) suggest that further patterns and structure are present, which could exploited in future work to obtain significant further speed-ups.

\section{$\Q_p$ and its implementation}\label{sec:padics}
In the field of computer algebra, it is common to obtain rational numbers not from floating-point real evaluations, which are prone to rounding errors, but instead from evaluations in a finite field $\mathbb{F}_p$ or $p$-adic field $\mathbb{Q}_p$.
The $p$-adic numbers $\Q_p$ are an extension of the rational numbers $\mathbb{Q}$ but are distinct from the real numbers~$\mathbb{R}$.
Ref.~\cite{DeLaurentis:2022otd} points out that $p$-adic fields are well suited to studying the singular limits of polynomials and rational functions, which is helpful for studying the partial-fractioning of rational functions as desired in this work.
Specifically, we will use the $p$-adic numbers to evaluate rational functions at special points that make chosen denominators ``small'' under a $p$-adic absolute value $|\cdot|_p$ (defined below), allowing the corresponding numerators to be isolated and examined, but not small under the standard absolute value $|\cdot|$, thereby ensuring that the evaluations are numerically stable. 
Introductory material on $p$-adic numbers and their use in computer algebra can be found in Ref.~\cite{zurGathenGerhard}, but here we will only mention that in this work we rely on the following facts:
\begin{enumerate}[beginpenalty=10000]
\item Given a prime number $p$, any rational number can be uniquely expanded as a $p$-adic series, i.e. a series $\sum_m c_m p^m$ with $c_m \in \{0, \ldots, p-1\}$. Such series are analogous to power series in a small parameter, so large powers of $p$ are ``small'' according to the so-called $p$-adic absolute value $|\cdot |_p$
\begin{equation}\label{eq:padic_metric}
\left|\frac{a}{b}p^n \right|_p = \frac{1}{p^n},
\end{equation}
where $a,$ $b,$ and $n$ are integers, and $a$ and $b$ are indivisible by $p$.
An example of such a series expansion is
\begin{equation}\label{eq:padic_half}
\frac{1}{2} = 4 + 3*7 + 3*7^2 + 3*7^3 + \bigO(7^4),
\end{equation}
where we have chosen to expand with $p=7$.
Equation~\eqref{eq:padic_half} can easily be verified by multiplying both sides of the equation by 2 and performing carries whenever $c_m \geq p$.
While the terms represented by the notation ``$\bigO(7^4)$'' on the RHS of eq.~\eqref{eq:padic_half} are clearly large under the usual absolute value, they are small under the so-called $7\textrm{-adic}$ absolute value defined by eq.~\eqref{eq:padic_metric} with $p=7$:
\begin{align}
\left|\frac{1}{2} - 4 - 3*7 - 3*7^2 - 3*7^3\right|_7 &= \left|-\frac{2401}{2}\right|_7 \nonumber \\
&= \frac{1}{7^4}.
\end{align}
\item Elementary arithmetical operations commute with performing $p$-adic expansions.
\item Any rational number is congruent (mod $p$) to the leading term of its $p$-adic expansion, assuming the denominator of that rational number is not a multiple of $p$. Taking the example in eq.~\eqref{eq:padic_half}, we have
\begin{equation}
\frac{1}{2} \equiv 4 \quad \textrm{(mod 7)},
\end{equation}
which can be verified by multiplying both sides by~2.
\end{enumerate}

There are many choices of practical methods to implement $p$-adic numbers on a computer. For example, just as with the real numbers, two common options are fixed-point and floating-point representations. While in this work we focus on reducing the number of required probes, which is independent of the choice of representation, let us mention that in practice we have here chosen to represent each $p$-adic number by an integer that is $p$-adically close to it, simply by truncating the $p$-adic series. For instance, the $p$-adic series on the RHS of eq.~\eqref{eq:padic_half} can be represented by the integer value 1201:
\begin{equation}\label{eq:truncate_series}
4 + 3*7 + 3*7^2 + 3*7^3 = 1201.
\end{equation}
Every $p$-adic evaluation in this work is performed as a rational evaluation at an integer point chosen in this way.
Unlike floating-point (real or $p$-adic) calculations, rational evaluations are exact and immune from rounding errors and loss of precision, regardless of whether they are performed as black-box evaluations on an analytic expression for $R$, as we do here, or as step-by-step solutions of a set of IBP equations, which is one of the applications envisaged.
In principle, rational evaluations should suffer from intermediate-expression swell, but we avoid this: in this work we surprisingly find that it is possible to use small primes $p \approx 101$ and series truncated to small powers of $p$, so that we typically perform evaluations at 10-digit integer values for the kinematic points and space-time dimension.\footnote{The number of digits varies between 4 and 20. In future work we intend to adopt a more sophisticated approach to reconstructing the spacetime-dimension variable, which we estimate would reduce the number of digits in all our larger integer evaluation points to around 10 digits.} Numerically solving IBPs and amplitudes at such integer points is well within the capabilities of standard publicly-available tools that internally use finite fields or exact integer arithmetic.

\section{Method}\label{sec:method}
In order to exploit the observations from sec.~\ref{sec:preliminary_remarks}, we have designed a method to reconstruct rational functions directly in partial-fractioned form eq.~\eqref{eq:R_pfed}, one partial-fractioned term $\frac{n_i}{d_i}$ at a time.
A set of all possible denominators $\{d_i\}$ is straight-forward to determine by examining the easily-obtainable (see sec.~\ref{sec:preliminary_remarks}) denominator of the common-denominator-form expression.
As explained in sec.~\ref{sec:preliminary_remarks}, the speedup in this paper will arise because for many of the possible denominators $d_i$, the corresponding $n_i$ is zero.
Reconstructing one partial-fractioned term at a time ensures that if a partial-fractioned term vanishes, we can notice this cheaply and avoid reconstructing its numerator.
A key further advantage of reconstructing one partial-fractioned term at a time is that our method will scale well for even larger rational functions than $R_*$, for reasons explained in sec.~\ref{subsec:reconstruction}.

Reconstructing one partial-fractioned term at a time also has other benefits, which we foresee but will leave to further work: for instance, noting that the bottleneck in cutting-edge calculations is sometimes a very small number of particularly large rational functions, we believe reconstructing one partial-fractioned term at a time would give maximum scope for on-the-fly observation of patterns that can be exploited in the remaining partial-fractioned terms. Examples of this might be the optimal choice of numerator variables for particular combinations of denominator factors, or the appearance of commonly-occurring integer or polynomial prefactors in the numerators of some partial-fractioned terms, or even (as we observe post-hoc in sec.~\ref{sec:results}) the appearance of identical numerators in several partial-fractioned terms.
Additionally, we believe our method of reconstructing one partial-fractioned term at a time will provide a powerful tool to better analytically understand---and eventually further exploit---the simplification that partial fractioning produces for rational functions in amplitudes and IBP expressions.

Let $R$ denote the rational function we wish to reconstruct and let $N$ be the number of variables it contains. We start by observing that if we can find a special $p$-adic point $\bar{x}$ at which the denominator $d_k$ of one partial-fractioned term $n_k / d_k$ becomes smaller than all the others, i.e. if for some $\bar{x} \in \Q_p^N$,
\begin{equation}\label{eq:pick_out_unique}
\exists k : \forall i \neq k, \quad |d_k (\bar{x})|_p < |d_i (\bar{x})|_p,
\end{equation}
then evaluating the complete rational function $R$ at that $p$-adic point $\bar{x}$ will give a series
\begin{equation}\label{eq:R_x}
R(\bar{x}) = \frac{n_k(\bar{x})}{d_k(\bar{x})} + \bigO(p^{-m+1}),
\end{equation}
where $m = -\log_p\left( |d_k(\bar{x})|_p \right)$.\footnote{In this work, $\log_p$ does not denote the $p$-adic logarithm sometimes seen in the mathematical literature, but instead just an ordinary logarithm with base $p$.}
In general, this series is $\bigO(p^{-m})$ and the coefficient of $p^{-m}$ gives useful information about $n_k(\bar{x})$.\footnote{The handling of rare exceptions to this is described in sec.~\ref{subsec:filtering} and sec.~\ref{subsec:reconstruction}.}
In particular, if $n_k = 0$, the $\bigO(p^{-m})$ term will vanish and so the leading term of the series $R(\bar{x})$ will be $\bigO(p^{-m+1})$ instead.
Furthermore, even when $n_k \neq 0$, we can use eq.~\eqref{eq:R_x} to obtain the leading $p$-adic digit of $n_k(\bar{x})$, in effect obtaining a finite-field evaluation of $n_k$.
By repeating for other values of $\bar{x}$ that still satisfy eq.~\eqref{eq:pick_out_unique} for the same $k$, we can gather sufficient information to reconstruct the analytic form of $n_k$, as we will explain below.
We have therefore devised a reconstruction strategy comprising several steps, which we will summarise here and then discuss in detail below.
\begin{enumerate}
\item Find the common-denominator-form denominator of $R$.
\item Enumerate a complete set of candidates for the denominators $\{d_i\}$ appearing in eq.~\eqref{eq:R_pfed}. At this stage, they are merely candidates, most of which will later turn out to have a vanishing numerator $n_k=0$.
\item Choose the $p$-adic evaluation points.
\item Filter the candidate denominators $\{d_i\}$ by using probes at the chosen points.
\item Reconstruct the numerator of one candidate by performing additional probes.
\item Repeat Steps 4 and 5 to reconstruct the other terms.
\end{enumerate}

From this terse summary the reader might see a superficial resemblance with the reconstruction procedure based on floating-point numbers described in Ref.~\cite{Laurentis:2019bjh}.
While both methods seek to produce compact results, we believe the techniques that we describe in our work offer a practical new capability to study and maximally exploit the simplification offered by partial fractioning, particularly in cases where the rational functions are very large or where floating-point rounding errors need to be avoided.

\subsubsection{Find the common denominator}
Step 1 is straight-forward, as explained in sec.~\ref{sec:preliminary_remarks} above.
The obtained common denominator $\Delta(x)$ is typically a product of polynomials $f_a(x)$, each raised to some power $\nu_a \in \mathbb{N}$:
\begin{equation}\label{eq:common_denominator}
\Delta(x) = \prod_a \left[f_a(x)\right]^{\nu_a}.
\end{equation}
In our example case, since $R_*$ is a function of 5-point massless kinematics, each polynomial $f_a(x)$ has degree~1 in the kinematic variables and space-time dimension. The code developed in this work therefore assumes linear factors, and we leave the extension to higher-order polynomial factors to future work.
Also, although not essential to what follows, let us mention that since $R_*$ comes from an IBP reduction into a pure basis of master integrals, each irreducible factor in the common denominator is a polynomial in either the kinematic variables or the space-time dimension $D$ but not both.
This feature helps us streamline some of the steps in our calculation.

\subsubsection{Enumerate candidates for $\{d_i\}$}\label{subsec:enumerate_candidates}
Having found the common denominator $\Delta(x)$, we now wish to enumerate a set of candidates for the denominators $\{d_i\}$ appearing in eq.~\eqref{eq:R_pfed}.
Each denominator $d_i$ will be the product of the elements of a subset $G_i$ of the factors $\{f_a\}$ in eq.~\eqref{eq:common_denominator}, each raised to some power $\mu_{i,a} \in \mathbb{N}$, i.e.
\begin{equation}\label{eq:di_factorised}
d_i(x) = \prod_b \left[f_b(x)\right]^{\mu_{i,b}},
\end{equation}
where the index $b$ is understood to run only over the factors present in
$d_i$.
It is convenient to implicitly define $\mu_{i,a}=0$ for the remaining factors $f_a$ not present in $G_i$.
Due to identities such as
\begin{equation}\label{eq:pf_basis_identities}
\frac{1}{yz(y+z)} + \frac{1}{y^2 (y+z)} - \frac{1}{y^2 z} \equiv 0,
\end{equation}
where $y$ and $z$ are generic variables, choices will need to be made about which partial-fractioned terms to use as a basis. 
The \textsc{MultivariateApart} package uses a basis determined solely by the common denominator and a specified variable ordering. Such a basis is helpful in analytic calculations because it aids vectorised addition of partial-fractioned expressions, but has a disadvantage in a numerical context because the individual partial fractioned terms $n_i/d_i$ in its output sometimes have higher powers $\mu_{i,b}$ of a denominator factor $f_b$ compared to the corresponding power $\nu_b$ in the common-denominator expression.
In this work we find it preferable\footnote{While this basis is helpful for reconstructing rational functions in partial-fractioned form from numerical evaluations, we do not claim it to necessarily be better for other purposes. Indeed, the size of $R_*$ as reconstructed in our basis is a few percent larger than in the basis of \textsc{MultivariateApart}.} to use a basis in which
\begin{equation}\label{eq:consistent_powers}
\mu_{i,b} \leq \nu_b \quad \forall i, b.
\end{equation}
Therefore, here in Step 2 we create an over-complete basis by enumerating a complete set $\{d_i\}$ of candidate denominators consistent with eq.~\eqref{eq:consistent_powers}, without paying regard to identities such as eq.~\eqref{eq:pf_basis_identities}.
Since in subsequent steps we perform the reconstruction one partial-fractioned term at a time, elements of the over-complete basis that are redundant due to equations like~\eqref{eq:pf_basis_identities} will automatically get filtered out after sufficiently many other basis terms have been reconstructed.
To create our over-complete basis, we use eq.~\eqref{eq:di_factorised} to tabulate every possible $d_i(x)$, requiring only that $G_i\ \subset \{f_a\}$ and that eq.~\eqref{eq:consistent_powers} holds.
Note that it is permissible for two distinct denominators $d_i(x), d_{i'}(x)$ to share the same factors (i.e. $G_i$=$G_{i'}$) and differ only in one or more of the powers $\mu_{i,b}, \mu_{i',b}$.

For reasons of efficiency, during Step 2 we discard any denominator $d_i$ if its factors $f_b \in G_i$ cannot all simultaneously vanish without making all dimensionful variables in $R$ vanish. Such denominators are not required in our basis since they can either be further partial fractioned, or would get filtered out by a later step.
For the same reason, we only allow denominators to contain at most 1 factor containing the space-time dimension variable.

\subsubsection{Choose $p$-adic evaluation points}\label{subsec:generate_weights}
Having generated a set of candidate denominators $\{d_i\}$, we next wish to perform $p$-adic probes of the black-box function $R$ in order to identify any denominator $d_i$ whose corresponding numerator $n_i$ vanishes.
Any $p$-adic evaluation point $\bar{x}$ will induce a weights vector
\begin{equation}\label{eq:weights}
w = \{-\log_p(|f_1(\overline{x})|_p),-\log_p(|f_2(\overline{x})|_p), \ldots\}.
\end{equation}
This weights vector $w$ largely determines which partial-fractioned term (or terms) will be picked out by a $p$-adic evaluation $R(\bar{x})$, since in general the contribution of a partial-fractioned term $n_i/d_i$ to the series~\eqref{eq:R_x} will be of $p$-adic order
\begin{equation}
-\sum_a \mu_{i,a} w_a,
\end{equation}
with some caveats to be described shortly.
Here $w_a$ is the component of $w$ corresponding to the factor $f_a$, i.e. $w_a = -\log_p(|f_a(\overline{x})|_p)$.

In this work we generate each $p$-adic evaluation point $\bar{x}$ by first selecting a desired weights vector $w$, in a manner described below, and then finding a point $\bar{x}$ satisfying eq.~\eqref{eq:weights} for that particular $w$.
In practice, eq.~\eqref{eq:R_x} may fail to hold if the partial-fractioned form of $R$ contains explicit factors of $p$, either in a denominator or in a term in one of the numerators.
It may also fail if by chance $n_i(\bar{x}) = 0 + \bigO(p)$ for some otherwise non-zero numerator $n_i$.
To avoid this, having selected $w$ we probe $R$ at 3 independent points $\bar{x}_1, \bar{x}_2, \bar{x}_3$, each satisfying eq.~\eqref{eq:weights} in a distinct $p$-adic field $\mathbb{Q}_{p_1}$, $\mathbb{Q}_{p_2}$, $\mathbb{Q}_{p_3}$ respectively.
In most cases, these 3 probes produce identical valuations $-\log_{p_i}(|R(\bar{x}_i)|_{p_i})$, and we will denote this valuation as~$R[w]$:
\begin{equation}
\begin{array}{rl}
R[w] &= -\log_{p_1}\left(|R(\bar{x}_1)|_{p_1}\right) \\
     &= -\log_{p_2}\left(|R(\bar{x}_2)|_{p_2}\right) \\
     &= -\log_{p_3}\left(|R(\bar{x}_3)|_{p_3}\right).
\end{array}
\end{equation}
If the 3 valuations are not identical, we adopt a heuristic procedure whereby we perform a few additional probes (each in a different $p$-adic field) until one valuation has been obtained in 3 more probes than all other valuations and in at least twice as many probes as all other valuations. In this work, we choose $p_n$ to be the $n$\textsuperscript{th} prime larger than 100, so $p_1=101, p_2 = 103, \ldots$.

It remains to be described how the weights are chosen. We generate a weights vector $w$ by first choosing an algebraically-independent\footnote{For the massless 5-point example $R_*$ in this paper, algebraic independence here is equivalent to linear independence, since all the factors $f_a$ are linear.} subset $\{h_b\}$ of the factors $\{f_a\}$ in eq.~\eqref{eq:common_denominator} and then assigning a weight $\tilde{w}_b$ to each factor $h_b$. The complete weights vector $w$ in eq.~\eqref{eq:weights} is inferred from $\{\tilde{w}_b\}$. For example, if $\{f_a\} = \{y, z, y+z\}$ and if we have assigned weights $\{\tilde{w}_b\} = \{5, 2\}$ to the factors $\{h_b\} = \{y, y+z\}$, i.e. $y=\bigO(p^5), y+z=\bigO(p^2)$, then it follows that $z=\bigO(p^2)$ and so $w = \{5, 2, 2\}$.

For comprehensiveness, we tabulated every possible subset of the factors $\{f_a\}$, although for efficiency we ignore factors containing (only) the space-time dimension variable. For each subset $\{h_b\}$, we generate weights in several ways which we have heuristically found to be useful:
\begin{itemize}
\item $\tilde{w}_b = 1 \quad \forall b$
\item for some $i$, set $\tilde{w}_b = 
\begin{cases}
4 \quad \textrm{if } b=i \\
1 \quad \textrm{if } b\neq i
\end{cases}
$
\item for some $i$, set $\tilde{w}_b = 
\begin{cases}
1 \quad \textrm{if } b=i \\
4 \quad \textrm{if } b\neq i
\end{cases}
$
\item $\{\tilde{w}_b\}$ is a permutation of $\{1, 2, \ldots, |\{\tilde{w}_b\}|\}$.
\end{itemize}
Note that some choices of $\{h_b\}$ and $\{\tilde{w}_b\}$ are equivalent to others. For example, the choice $\{y, z\} = \{\bigO(p), \bigO(p^4)\}$ leads to the same $w$ as the choice $\{y+z, z\} = \{\bigO(p),\bigO(p^4)\}$.
We highlight that the above-generated weight vectors, despite only containing small entries $\tilde{w}_b \le 4$, are sufficient to uniquely pick out individual partial-fractioned terms in the manner that we will describe in sec.~\ref{subsec:reconstruction}.

In total we obtain 6,062 distinct weight vectors, which we will use in the remaining steps to generate $p$-adic probe points for both filtering and reconstruction purposes.

\subsubsection{Filter candidate denominators} \label{subsec:filtering}
We now wish to filter the candidates generated in Step~\ref{subsec:enumerate_candidates}.
To do so, we perform a total of 6,062 $p$-adic probes per prime field, each probe corresponding to one of the weight vectors $w$ generated in the previous step.
We expect it to be possible to reduce the number of probes by being more selective about the choices of $\{h_b\}$ and $\{w_b\}$, but since this number is already relatively small we leave that to future work and here simply state that these probes are sufficient.
Note that although Step~\ref{subsec:repeat} requires us to repeat the filtering step many times, these probes do not need to be repeated if the results are cached and the probes are performed with exact integer arithmetic (as in our code) or with sufficiently high $p$-adic precision in a floating-point-style $p$-adic evaluation code.

Having performed the probes in the manner described above, we can now filter the candidate denominators $\{d_i\}$.
We do this by discarding any candidate $d_i$ if there exists a $w$ for which both of the following relations hold:
\begin{equation}\label{eq:filtering}
R[w] > - \sum_b \mu_{i,b} w_b
\end{equation}
\begin{equation}\label{eq:when_to_filter}
\left<f_b : \mu_{i,b} \neq 0\right> = \left< f_{b'} : \left(\mu_{i,b'} \neq 0 | w_{b'} \neq 0\right) \right>
\end{equation}
where the notation $\left< \quad \right>$ indicates the polynomial ideal (see e.g. Ref.~\cite{zurGathenGerhard} for a definition) and $|$ denotes the Boolean~\textsc{or} operation.
Here equation~\eqref{eq:filtering} indicates that the divergence of $R$ at points with $p$-adic weight $w$ is milder than it would have been if a term with denominator $d_i$ had been present, assuming $n_i$ does not vanish at the probed points.
Equation~\eqref{eq:when_to_filter} is necessary to ensure that indeed $n_i$ does not vanish at the probed points; for instance it is important that a partial-fractioned term like $\frac{x^2}{y^4 z^5}$ does not get filtered out by probing at $\{x, y, z\} = \{\bigO(p), \bigO(p), \bigO(p) \}$ and obtaining a series $\bigO(p^{-7})$, which is less divergent than the $\bigO(p^{-9})$ series that one might have expected based on the behaviour of the denominator $y^4 z^5$ alone.

\subsubsection{Reconstruct a partial-fractioned term}\label{subsec:reconstruction}
Having performed the filtering, we are now ready to reconstruct a partial-fractioned term.
We remind the reader that in this work we seek to reconstruct rational functions one partial-fractioned term $n_k/d_k$ at a time, to obtain the benefits described at the start of this section.
This requires finding a probe point $\bar{x}$ satisfying eq.~\eqref{eq:pick_out_unique}.
At first sight, it might be expected that this could potentially require exponentially-large values for the weights $w_b$ that generate $\bar{x}$, which in turn would lead to a need for $p$-adic evaluations at very high $p$-adic precision.
Remarkably, however, we find in this work (albeit without proof) that it is always possible to generate a suitable weight vector $w$ whose weights $w_b$ are small and for which any point $\bar{x}$ satisfying eq.~\eqref{eq:weights} will also satisfy eq.~\eqref{eq:pick_out_unique}.
In fact, preliminary work indicates that in future calculations we will continue to be able to pick out one partial-fractioned term at a time while keeping the weights $w_b$ modest in size.

The weight vectors we use for reconstruction are essentially based on those generated in sec.~\ref{subsec:generate_weights}, which we highlight are small: $w_b$ never exceeds 4.
We remind the reader that for $R_*$ the factors of the common denominator $\Delta$ contain either $D$ or the kinematic factors but not both, and that the weight vectors in sec.~\ref{subsec:generate_weights} had ignored factors containing $D$.
We find that there always exists a weight vector amongst those generated in sec.~\ref{subsec:generate_weights} that selects either a unique partial-fractioned term, in accordance with eq.~\eqref{eq:pick_out_unique}, or several partial-fractioned terms whose denominators differ only by a $D$-dependent factor.
To ensure we reconstruct a single partial-fractioned term at a time, having identified such a weight vector we multiply all its entries $w_b$ by a heuristically-chosen factor of 2 or 3 as necessary, and then set one of the $D$-related entries to 1.\footnote{A simpler alternative would be to treat the $D$-dependent factors on an equal footing to the kinematic factors in all filtering and reconstruction steps. This may lead to a smaller maximum size for $w_b$, and hence simpler $p$-adic evaluation points.}
The weights vector $w$ obtained in this way uniquely picks out a single partial-fractioned term as desired.

Next, having obtained one such $w$ and the denominator $d_k$ that it uniquely picks out, we construct a simple fully-generic polynomial ansatz for $n_k$, constrained only by its mass dimension and the number of variables.\footnote{Note that in this work, we do not follow the common practice of ``setting a variable to 1'', as this would remove our capability to make that variable $p$-adically small when performing probes. Instead, we always work with homogeneous polynomials of well-defined mass dimension. We emphasise that this does not add any inefficiency, since a homogeneous polynomial of fixed degree has the same number of free parameters as an inhomogeneous polynomial with 1 fewer variable but the same maximum degree.}
The mass dimension of $n_k$ can be deduced from the mass dimensions of $R$ and $d_k$.
We determine the number of variables by observing that in general when performing partial fractioning, the numerator $n_k$ of a partial-fractioned term is uniquely defined only up to the addition of arbitrary polynomial multiples of one or more of the factors $f_b \in G_k$ of $d_k$.
Since in our case we have only linear polynomials as factors, the redundancy in $n_k$ can be removed by using each linearly-independent factor in $d_k$ to eliminate one variable from $n_k$. The ansatz that we construct for $n_k$ therefore contains only the residual variables.

Now given $w$, $d_k$, and an ansatz for $n_k$, we generate as many points $\bar{x}$ as there are free parameters in the ansatz, which in this work turns out to be up to 56.
Each of these points $\bar{x}$ is chosen to satisfy eq.~\eqref{eq:weights}, for a fixed prime $p_1=101$, and at each point we evaluate $R(\bar{x})$ to obtain a sample $n_k(\bar{x})$ modulo $p_1$ according to eq.~\eqref{eq:R_x}.
By matrix inversion we obtain an analytic form for $n_k$ modulo $p_1$, which we then verify by generating a few more points while keeping $p_1$ fixed.
Note that standard polynomial interpolation methods~\cite{zurGathenGerhard} scale linearly or quadratically with the number of unknowns whereas the ansatz-based interpolation techniques often seen in particle physics scale cubicly due to the cost of matrix inversion.
We emphasise that reconstructing one partial-fractioned term at a time offers the key advantage that the matrix to be inverted is small (no larger than 56-by-56 in this work), which means our approach can scale well in future calculations.

Having obtained the analytic form of $n_k$ modulo $p_1$, we then repeat for a few more primes $p_2=103, p_3=107, \ldots$ until we have sufficient information to reconstruct the exact rational form of $n_k$ using the Chinese remainder theorem and rational number reconstruction~\cite{10.5555/270146}.
Note that in contrast to finite-field calculations where machine-word-sized primes $p \sim 4*10^9$ or $p \sim 2*10^{19}$ are standard, $p$-adic calculations can benefit from faster arithmetic using small primes such as $p=101$ (or even smaller), regardless of whether the probes are performed with exact integers, floating-point, fixed-point, or other representations.
Since the coefficients in partial-fractioned form are simpler than in common-denominator form, we typically require only 3 or 4 primes despite the small size of the primes, plus one more prime for performing checks.
The number of required prime fields required to reconstruct a free parameter is proportional to the complexity of that free parameter\footnote{For simplicity, in this work we used the same number of prime fields for all the free parameters in a given numerator $n_i$, but optimising by removing this redundancy would be straight-forward.} and so in a few cases we used up to 9 prime fields (plus one more for checks).

\subsubsection{Repeat to reconstruct more terms}\label{subsec:repeat}
Having reconstructed a term $R_{rec} = n_k/d_k$, we next wish to reconstruct the rest of the terms.
To do so, we repeat the filtering (sec.~\ref{subsec:filtering}) and reconstruction (sec.~\ref{subsec:reconstruction}) steps, but now seek to reconstruct $R - R_{rec}$.
All probes are hence performed on $R - R_{rec}$ instead of $R$, and in practice we do this by separately probing $R$ and $R_{rec}$ and then subtracting.
Whenever we obtain a new reconstructed term, we then add it onto $R_{rec}$ and then repeat the filtering and reconstruction steps again.

As a technical point, we should highlight the importance of lifting each numerator $n_k$ from $\F_p$ to the rationals in the previous step \emph{before} attempting to reconstruct further terms.
If $n_k/d_k$ is added onto $R_{rec}$ after merely reconstructing in $\F_p$, subleading $p$-adic contributions from $n_k/d_k$ will continue to contribute to evaluations of $R - R_{rec}$ and so prevent the reconstruction of other terms with denominators similar to $d_k$.

Regarding efficiency, we note that $R_{rec}$ is cheap to probe, since we always have an analytic expression for it.
Probes of $R$ are therefore the main cost that we seek to minimise. Note that as mentioned in sec.~\ref{subsec:filtering}, the probe points used in the filtering step are fixed and so if the probes are exact and the value of $R$ at each of those points is saved once, there is no need to probe $R$ again when repeating the filtering step.\footnote{A cache of this sort should be straight-forward to implement in software, although for simplicity in this proof-of-concept work we have not done so.}
The reconstruction step, conversely, requires probing $R$ (and $R_{rec}$) at points specially chosen to select a specific denominator.
In principle, large efficiency gains could be obtained in the reconstruction step by recycling probes performed in previous iterations of the reconstruction step for similar denominators, but this would require careful further study and so we leave it to future work.

\section{Results and Discussion}\label{sec:results}

By employing the technique presented in sec.~\ref{sec:method}, we reconstructed $R_*$ in full. We did so using no knowledge of $R_*$ other than its mass dimension, its common denominator, and the results of ``black-box'' $p$-adic probes. As explained above, the $p$-adic probes were used for two purposes: filtering (see sec.~\ref{subsec:filtering}) and reconstruction (see sec.~\ref{subsec:reconstruction}). For filtering we performed 6,023 $p$-adic probes per prime field, while for reconstruction we performed a total of $6 \times 10^4$ probes per prime field.
As mentioned above, we use very small-sized primes $p = \bigO(100)$ and yet we mostly require only 3 or 4 prime fields (plus one more for checks) thanks to the relative simplicity of numerical coefficients in partial-fractioned form compared to common-denominator form.
We verified that the reconstructed expression is exactly equal to the original expression.
As shown in Table.~\ref{tab:results}, the reconstructed result is 134 times smaller than the common-denominator form that would be produced by conventional finite-field methods.
The total number of free parameters fitted was 52,527.
Each free parameter accounts for 1 of the $6\times 10^4$ probes (per prime field) performed for reconstruction, the remaining probes being used for checking purposes out of prudence, although these checks never failed.
We remind the reader that we never fit more than 56 free parameters at a time since, as discussed extensively in the previous sections, our result was reconstructed one partial-fractioned term at a time, bringing the many advantages already described.

\begin{table}
\caption{
\label{tab:results}
Comparison of original and reconstructed form of $R_*$.
Original expression is in common-denominator form, with numerator fully expanded and denominator fully factorised.
Sizes are as reported using \texttt{ByteCount} in \textsc{Mathematica}.
Number of free parameters is obtained by counting the number of terms in the fully-expanded numerator(s). 
}
\begin{tabular}{c|c|c}
Expression	&	Size	& Parameters to fit \\
\hline
Original & 605 MB &	1,369,559 \\
Reconstructed	& 4.5 MB	& 52,527 (\emph{of which 15,403 non-zero})
\end{tabular}
\end{table}

The reconstructed result exhibits further structure which in future work it would be beneficial to study and exploit.
This is best seen and discussed with an example.
We will consider the following reconstructed terms
\begin{multline}\label{eq:some_reconstructed_terms}
\frac{\frac{45}{1024} s_{45}^6 s_{12}^3}{(D-3) s_{34}^4 s_{51} (-s_{23}+s_{45}+s_{51})^3}\\
+\frac{\frac{9}{5120} s_{45}^6 s_{12}^3}{(D-1) s_{34}^4 s_{51} (-s_{23}+s_{45}+s_{51})^3}\\
-\frac{\frac{693}{5120} s_{45}^6 s_{12}^3}{(2 D-7) s_{34}^4 s_{51} (-s_{23}+s_{45}+s_{51})^3}\\
-\frac{\frac{3}{1024} s_{45}^6 s_{12}^3}{s_{34}^4 s_{51} (-s_{23}+s_{45}+s_{51})^3}\\
+\frac{-\frac{45 s_{45}^6 s_{51}^2}{1024}-\frac{135 s_{45}^6 s_{51} s_{12}}{1024}-\frac{135 s_{45}^6 s_{12}^2}{1024}}{(D-3) s_{34}^4 (s_{23}-s_{45}-s_{51})^3}\\
+\frac{-\frac{9 s_{45}^6 s_{51}^2}{5120}-\frac{27 s_{45}^6 s_{51} s_{12}}{5120}-\frac{27 s_{45}^6 s_{12}^2}{5120}}{(D-1) s_{34}^4 (s_{23}-s_{45}-s_{51})^3}\\
+\frac{\frac{693 s_{45}^6 s_{51}^2}{5120}+\frac{2079 s_{45}^6 s_{51} s_{12}}{5120}+\frac{2079 s_{45}^6 s_{12}^2}{5120}}{(2 D-7) s_{34}^4 (s_{23}-s_{45}-s_{51})^3}\\
+\frac{-\frac{3 s_{45}^6 s_{51}^2}{1024}-\frac{9 s_{45}^6 s_{51} s_{12}}{1024}-\frac{9 s_{45}^6 s_{12}^2}{1024}}{s_{34}^4 (-s_{23}+s_{45}+s_{51})^3},
\end{multline}
which form a small part of our full reconstructed result.
Here $s_{ij}$ are the 5 kinematic variables of $R$.

Firstly, let us note that 70\% of the free parameters that we fitted turn out to be zero, as anticipated from the discussion at the end of sec.~\ref{sec:preliminary_remarks}.
This is can be seen in expression~\eqref{eq:some_reconstructed_terms} in the following way. Expression~\eqref{eq:some_reconstructed_terms} contains 16 numerator terms and therefore accounts for 16 of the 15,403 non-zero free parameters mentioned in Table~\ref{tab:results}.
Considering the first term in~\eqref{eq:some_reconstructed_terms}, we note that a priori there was no reason for the numerator to only contain a term $\sim s_{45}^6 s_{12}^3$; it could equally well have contained other mass-squared-dimension-9 combinations of $s_{45}$ and $s_{12}$, such as $s_{45}^2 s_{12}^7$.
To obtain~\eqref{eq:some_reconstructed_terms} we therefore actually fitted a total of 220 free parameters, of which 204 turned out to be zero.
These terms thus account for 220 of the 52,527 free parameters mentioned in Table~\ref{tab:results}.
Identifying the vanishing parameters in advance would reduce the number of parameters to be fitted from 52,527 to 15,403, and reduce the number of probes correspondingly.

Secondly, some of the numerators in our reconstructed result are linearly related to each other by a simple integer multiple.
Looking at two of the numerators in the expression~\eqref{eq:some_reconstructed_terms}
\begin{equation}
n_1 = - \frac{45 s_{45}^6 s_{51}^2}{1024}-\frac{135 s_{45}^6 s_{51} s_{12}}{1024}-\frac{135 s_{45}^6 s_{12}^2}{1024},
\end{equation}
\begin{equation}
n_2 = - \frac{9 s_{45}^6 s_{51}^2}{5120}-\frac{27 s_{45}^6 s_{51} s_{12}}{5120}-\frac{27 s_{45}^6 s_{12}^2}{5120},
\end{equation}
we notice $n_1 = 25 n_2$.
If this can economically be discovered prior to reconstruction, it would further reduce the number of free parameters to be fitted and thus the number of probes required.

Thirdly, we notice that in some cases it is possible to combine several of our reconstructed terms and obtain a simpler expression. For example, if we combine together all the terms in expression~\eqref{eq:some_reconstructed_terms}, we obtain the following simple term:
\begin{equation}\label{eq:reconstructed_terms_combined}
-\frac{\frac{3}{512} D \left(D^2-4\right) s_{45}^6 (s_{51}+s_{12})^3}{(D-3) (D-1) (2 D-7) s_{34}^4 s_{51} (-s_{23}+s_{45}+s_{51})^3}.
\end{equation}
Note however that the first two properties do not necessarily imply the third, and we observe from examining other reconstructed terms that combining them in this manner does not always simplify them.
The results in Table~\ref{tab:results} do not employ any such recombination of terms, and further study is required to understand which cases are amenable to such simplification and to devise a manner to exploit it during the reconstruction itself, rather than afterwards.
This is an interesting direction for exploration, with the potential to yield a further order-of-magnitude reduction in the number of free parameters to be fitted, the number of probes required per prime field, and the size of the final result.
Additionally, since the numerical coefficient $\frac{3}{512}$ in~\eqref{eq:reconstructed_terms_combined} is somewhat simpler than coefficients like $\frac{2079}{5020}$ in~\eqref{eq:some_reconstructed_terms}, fewer prime fields would be required to fit this coefficient.

It is worthwhile to note that while the patterns and structure exploited in this work---as well as those left for future work---could be studied post-hoc by partial-fractioning an expression obtained by conventional means, our method of reconstructing one partial-fractioned term at a time provides the capability to study and exploit these structures \emph{during} the reconstruction process.
For cutting-edge calculations where obtaining any analytic expression in the first place is the principal challenge and goal, we believe this new capability will be a valuable asset.

Going further, we emphasise the desirability of analytically studying the simplifications explored in this work, possibly in conjunction with the observations in Refs~\cite{Laurentis:2019bjh,DeLaurentis:2022otd,Campbell:2022qpq}.
We hope our techniques will prove to be useful tools in this regard, with benefits for our theoretical understanding as well as the speed of calculations.

\section{Conclusion and outlook}\label{sec:conclusion}
In this work we presented a technique to reconstruct rational functions in partial-fractioned form by using numerical evaluations in $p$-adic fields.
Using the example of $R_*$, a highly non-trivial rational function at the edge of the reach of current calculational techniques, we showed that our technique can reconstruct such functions using 25 times fewer numerical probes than conventional techniques based on finite-field probes, and yields a 100-fold simplification in the size of the reconstructed result.

The rational functions appearing in the final results for amplitudes are typically simpler than the IBP reductions used to compute them, and so having demonstrated our techniques on this large expression from the 2-loop 5-point massless non-planar IBP table, we are confident that our techniques can efficiently reconstruct any 2-loop 5-point massless full-colour amplitude.
The natural next step would be to apply our techniques to 2-loop 5-point processes with masses, most of which currently remain unknown.
We expect our partial-fractioned reconstruction technique will continue to be beneficial in the massive case, although some technical changes will be required in our code since we can no longer rely on the denominator factors $f_a$ to be linear.
More generally, experience shows that partial-fractioning produces simplifications in amplitudes whenever several kinematic scales are present, and so we expect our techniques will be applicable to a wide range of higher-point or higher-loop amplitudes.

Since rational function reconstruction can be used in a wide variety of multi-loop contexts, ranging from integrand construction to IBPs to differential equations, we focussed in this work on reducing the number of probes without regard to the choice of method for performing the probes themselves, for which many options exist.
One option is a floating-point-like representation of $p$-adic numbers, which we believe can be achieved with relatively little overhead compared to finite field probes particularly if small primes $p\sim\bigO(100)$ are used, but may be susceptible to loss of $p$-adic precision depending on the specific framework used for performing the probes.
Another option is to use exact integer probes at points $p$-adically close to the desired $p$-adic probe points---see eq.~\eqref{eq:truncate_series}.
Such probes completely avoid precision loss and would benefit from a speedup by using small primes but at present it may still be prudent to assume each probe to be slower than a conventional finite-field probe.
Since integer probes are exact, we expect them to be well suited to reconstruction probe recycling: using the same probes to reconstruct numerators for multiple terms that have similar denominators.
This would further reduce the number of required probes beyond the improvements already reported here and could partly or fully mitigate against any additional cost of $p$-adic probes compared to finite-field probes.

Finally, we observed that our reconstructed result for $R_*$ displays further patterns and structures which would be worthwhile to study, understand, and exploit in future work.
These observations provide hints of the potential to obtain even further improvements in the speed and reach of this calculational method, as well as potential avenues for starting to seek further understanding of the physical origin of these simplifications and of the structure of the rational functions appearing in scattering amplitudes and IBPs.

\begin{acknowledgments}
We are grateful to Federico Buccioni, Fabrizio Caola, Fernando Febres Cordero, Stephen Jones, Stefano Laporta, Andrew M\textsuperscript{c}Clement, and Alex Mitov for helpful discussions. We also thank the authors of Ref.~\cite{Agarwal:2021vdh} for providing analytic expressions for the IBP coefficients used as examples in this work.
This work has been funded by the European Research Council (ERC) under the European Union's Horizon 2020 research and innovation programme (grant agreement no. 804394 \textsc{hipQCD}).
The work was also supported in part by the U.S. Department of Energy under grant DE-SC0010102.
We are also grateful to the Galileo Galilei Institute for hospitality and support during the scientific program on ``Theory Challenges in the Precision Era of the Large Hadron Collider'', where part of this article was written.

\end{acknowledgments}

\bibliography{p-adic_reconstruction.bib}% Produces the bibliography via BibTeX.

\end{document}